\documentclass[twoside,a4paper,final]{aipproc}

\layoutstyle{6x9}

\usepackage{amsmath}
\usepackage{amssymb}
\usepackage{bbm}
\usepackage{array}
\renewcommand{\slash}[1]{\ensuremath{ #1 \mspace{-8mu} / }}

\begin{document}

\title{Impact of Four-Quark Condensates on In-Medium Effects of Hadrons\footnote{The work is supported by BMBF and GSI.}}

\author{R. Thomas}{
   address={Forschungszentrum Dresden-Rossendorf, PF 510119, 01314 Dresden, Germany}
}

\author{T. Hilger}{
   address={Institut f\"ur Theoretische Physik, TU Dresden, 01062 Dresden, Germany}
}

\author{S. Zschocke}{
   address={TU Dresden, Lohrmann-Observatorium, 01062 Dresden, Germany}
}

\author{B. K\"ampfer}{
   address={Forschungszentrum Dresden-Rossendorf, PF 510119, 01314 Dresden, Germany},
   altaddress={Institut f\"ur Theoretische Physik, TU Dresden, 01062 Dresden, Germany}
}



\begin{abstract}
Spectral properties of hadrons in nuclear matter are treated in the framework of QCD sum rules. The influence of the ambient strongly interacting medium is encoded in various condensates. Especially, the structure of different four-quark condensates and their density dependencies in light quark systems are exemplified for the $\omega$ meson and the nucleon.
\end{abstract}

\maketitle

\section{Introduction}
Strongly interacting matter, being subject of QCD, appears in different phases, depending on the temperature and the chemical potential, which characterize a thermalized system. Thereby, distinct regions of the corresponding QCD phase diagram can exhibit rich structures, especially in the areas of high temperature and density, where deconfinement is expected. However, also for moderate temperature and density in the hadronic phase one expects a change in properties of matter and its constituents, the hadrons. In photoproduction off nuclei (CB-TAPS~\cite{Trnka:2005ey}, TAGX~\cite{Huber:2003pu}, LEPS~\cite{Ahn:2004id}, CLAS~\cite{pk:Weygand2005}) or by measurements of the dilepton channel in $\mathit{C}+\mathit{C}$ collisions (HADES~\cite{temp-pk:HADES}), $p+A$ reactions (KEK~\cite{Naruki:2005kd}) or heavy-ion collisions (e.g.\ CERES~\cite{Adamova:2002kf}, NA60~\cite{Damjanovic:2005ni}, STAR~\cite{Adams:2003cc}) the study of medium-induced modifications for light mesons is pursued.
Since hadrons are confined composite objects of the fundamental degrees of freedom of QCD, excited from a ground state, it becomes possible to probe changes of the QCD vacuum at various conditions. A quantitative description of hadron properties can be linked, within the QCD sum rule approach, to expectation values of quark and/or gluon operators --- the QCD condensates. We consider here the four-quark condensates and relate their density dependencies to spectral modifications of $\omega$ meson and nucleon in cold nuclear matter.

\section{Four-Quark Condensates in QCD Sum Rules}
\vspace{-0.1cm}
QCD sum rules~\cite{Shifman:1978bx} match non-perturbative QCD condensates to hadronic quantities. Therefore each hadron considered is represented by an interpolating current $j_\mu$ built out of quarks and gluons. The correlation function
\vspace{-0.15cm}\begin{equation}
\Pi_{\mu \nu} (q) = i \int d^4 x \, e^{iqx} \, \langle \Psi | T[j_\mu (x) j_\nu (0)] | \Psi \rangle
\end{equation}\vspace{-0.45cm}

\noindent is evaluated, on the one side for time-like momenta $q$, where the physical hadrons are realized, and on the other side in terms of quarks and gluons via an operator product expansion (OPE) for large space-like euclidian momenta. Both representations for $\Pi$ are related using subtracted dispersion relations. Thus, integrals over spectral densities on the hadronic side correspond to QCD condensates, which enter as expectation values of the OPE terms. The leading condensates are the chiral condensate $m_q \langle \bar{q} q \rangle$, a measure for chiral symmetry breaking, the gluon condensate $\langle \tfrac{\alpha_s}{\pi} G^2 \rangle$,  related by the trace anomaly to a breaking of scale invariance, the mixed quark-gluon condensate $\langle \bar{q} g_s \sigma G q \rangle$, and condensates of mass dimension 6: the triple gluon condensate $\langle g_s^3 G^3 \rangle$ and four-quark condensates $\langle \bar{q} \Gamma q \bar{q} \Gamma' q \rangle$ ($\Gamma$ is symbolic for all possible structures specified below).
The extension to finite temperatures and densities induces on the OPE side modification of the condensates. For cold nuclear matter, modelled in leading order as Fermi gas of non-interacting nucleons, this dependence
$ \langle \mathcal{O} \rangle = \langle \mathcal{O} \rangle_0 + \tfrac{n}{2M_N} \langle \mathcal{O} \rangle_N $
is linear in the baryon density $n$ and dictated by nucleon matrix elements $\langle \mathcal{O} \rangle_N \equiv \langle N | \mathcal{O} | N \rangle$.

\vspace{2mm}
\begin{table}[h]
\begin{tabular*}{14.7cm}{m{2.2cm}|m{2.95cm}m{0.15cm}m{3.2cm}|m{4.0cm}}
\cline{1-2}\cline{4-5}
$\langle \bar{q} q \bar{q} q \rangle$ & $\langle \bar{q} \lambda^A q \bar{q} \lambda^A q \rangle$ && $\langle \bar{q} \slash{v} q \bar{q} \slash{v} q /v^2 \rangle$ & $\langle \bar{q} \slash{v} \lambda^A q \bar{q} \slash{v} \lambda^A q /v^2 \rangle$\\
$\langle \bar{q} \gamma_\alpha q \bar{q} \gamma^\alpha q \rangle$ & $\langle \bar{q} \gamma_\alpha \lambda^A q \bar{q} \gamma^\alpha \lambda^A q \rangle$ && $\langle \bar{q} \sigma_{\alpha \beta} v^\beta q \bar{q} \sigma^{\alpha \gamma} v_\gamma q /v^2 \rangle$ & $\langle \bar{q} \sigma_{\alpha \beta} v^\beta \lambda^A q \bar{q} \sigma^{\alpha \gamma} v_\gamma \lambda^A q /v^2 \rangle$ \\
$\langle \bar{q} \sigma_{\alpha \beta} q \bar{q} \sigma^{\alpha \beta}  q \rangle$ & $\langle \bar{q} \sigma_{\alpha \beta} \lambda^A q \bar{q} \sigma^{\alpha \beta}  \lambda^A q \rangle$ && $\langle \bar{q} \gamma_5 \slash{v} q \bar{q} \gamma_5 \slash{v} q /v^2 \rangle$ & $\langle \bar{q} \gamma_5 \slash{v} \lambda^A q \bar{q} \gamma_5 \slash{v} \lambda^A q /v^2 \rangle$ \\
$\langle \bar{q} \gamma_5 \gamma_\alpha q \bar{q} \gamma_5 \gamma^\alpha q \rangle$ & $\langle \bar{q} \gamma_5 \gamma_\alpha \lambda^A q \bar{q} \gamma_5 \gamma^\alpha \lambda^A q \rangle$ && $\langle \bar{q} \slash{v} q \bar{q} q \rangle$ & $\langle \bar{q} \slash{v} \lambda^A q \bar{q} \lambda^A q \rangle$ \\
$\langle \bar{q} \gamma_5 q \bar{q} \gamma_5 q \rangle$ & $\langle \bar{q} \gamma_5 \lambda^A q \bar{q} \gamma_5 \lambda^A q \rangle$ && $\langle \bar{q} \gamma_5 \gamma_\alpha q \bar{q} \gamma_5 \sigma^{\alpha \beta} v_\beta q \rangle$ & $\langle \bar{q} \gamma_5 \gamma_\alpha \lambda^A q \bar{q} \gamma_5 \sigma^{\alpha \beta} v_\beta \lambda^A q \rangle$ \\
\cline{1-2}\cline{4-5}
\end{tabular*}
\caption{Left part: List of all four-quark condensates in vacuum for one flavor $q$ ($\Gamma \in \{\mathbbm{1}, \gamma_\alpha, \sigma_{\alpha \beta}, \gamma_5 \gamma_\alpha, \gamma_5 \}$ are elements of the Clifford algebra, $\lambda^A$ Gell-Mann matrices). The columns correspond to two color singlet structures which can, for identical flavor, be transformed into each other by Fierz relations. Right part: The four-quark condensates which arise additionally in medium.\newline}
\label{tab:fqclist}
\end{table}
\vspace{2mm}

For the following discussion it is important to distinguish the different four-quark condensate structures. A list of all possible four-quark condensates follows from the demanded Lorentz invariance, invariance with respect to parity and time reversal and symmetry under $SU(3)_{color}$ of the normal ordered expectation values $\langle \Psi | : \bar{q} \Gamma q \bar{q} \Gamma' q : | \Psi \rangle$. Flavor symmetry will further reduce the number of independent four-quark condensates. The left part of Tab.~\ref{tab:fqclist} shows these condensates for $n_f = 1$ flavors in vacuum. In a medium with four-velocity $v_\mu$ they become density dependent and additional Lorentz scalars (right part of Tab.~\ref{tab:fqclist}) can be formed giving rise to additional structures.
In each part of Tab.~\ref{tab:fqclist} the two possible color singlets are listed which transform into each other by Fierz relations in the case of identical flavors presented here. Note that this four-quark condensate catalog increases when considered for more flavor degrees of freedom, also because flavor mixing terms are realized.

In particular sum rule calculations, distinct linear combinations of four-quark condensates occur and their relevance will be case-specific. This prevents from an extraction of a particular four-quark condensate although the QCD condensates are generic for the whole hadronic spectrum. For quantitative evaluations the four-quark condensates, even in vacuum not well-known, are often expressed by the squared chiral condensate, i.e.\ a factorization ansatz motivated by large-$N_c$ arguments~\cite{Leupold:2005eq}. In linearized form w.r.t.\ the density, $\langle \bar{q} q \rangle^2$ provides a first guess for the four-quark condensates and their density dependencies. Deviations of this ansatz are parameterized as
$\langle \bar{q}_{\mathrm f_1} \Gamma_1 q_{\mathrm f_1} \bar{q}_{\mathrm f_2} \Gamma_2 q_{\mathrm f_2} \rangle = A \kappa^\mathrm{vac} + B \kappa^\mathrm{med} n $,
where $A$, $B$ are individual constants for each structure $\Gamma_{1,2}$ tabulated (and generalized to flavors $f_{1,2}$; $B$ contains also the density dependence of $\langle \bar{q} q \rangle$) and $\kappa^\mathrm{med}$ allows different assumptions for the behavior at finite density (e.g.\ $\kappa^\mathrm{med} = 1$ is the factorization limit, $\kappa^\mathrm{med}=0$ means no density dependence at all).

\vspace{0.3cm}
\begin{figure}[htb]
\includegraphics[width=5.8cm,angle=270]{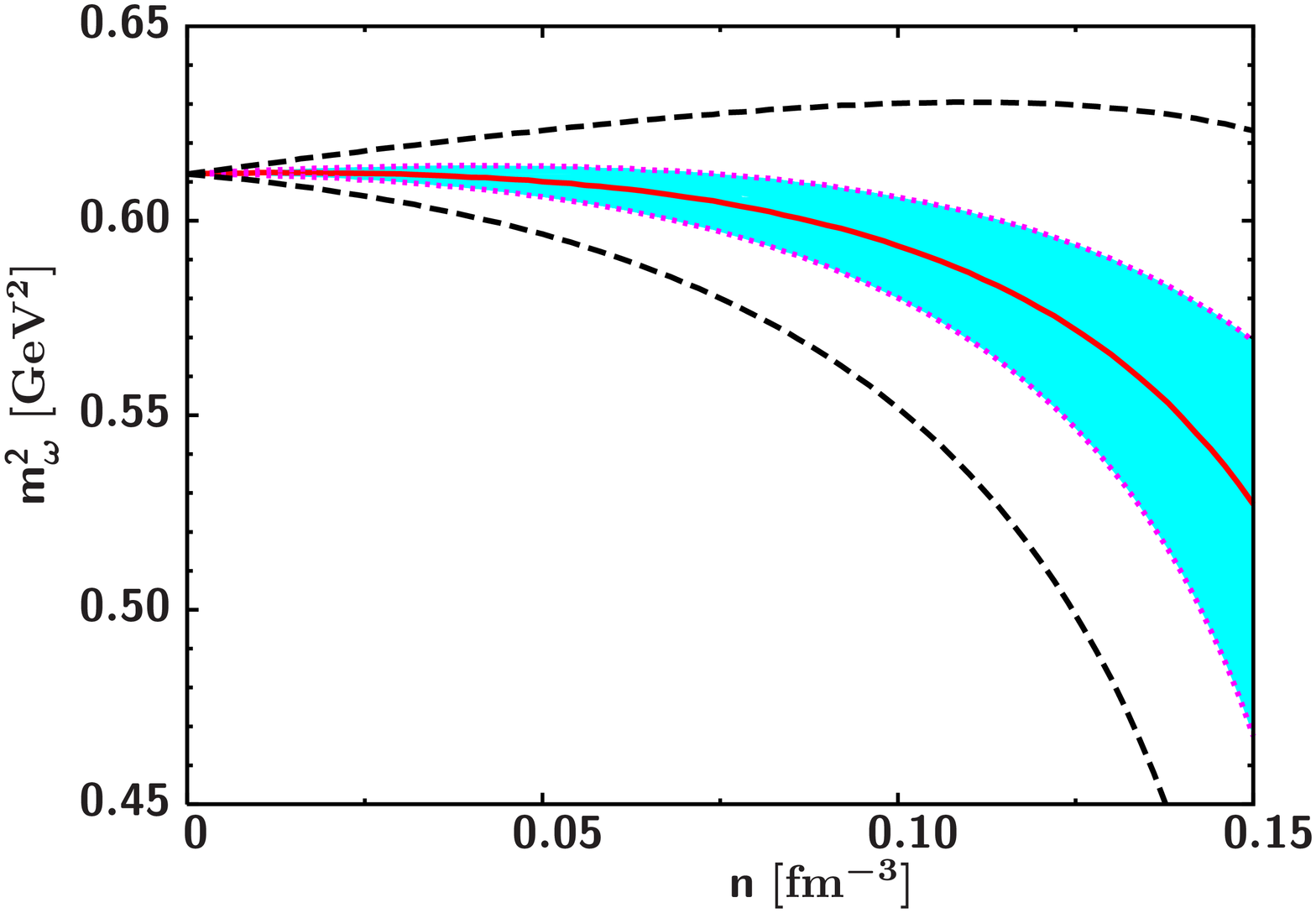}
\caption{The $\omega$ meson mass parameter $m_\omega^2$ as a function of the baryon density $n$ from a full QCD sum rule evaluation~\cite{Thomas:2005dc} for $\kappa^\mathrm{med}_\omega = 4$ and condensates up to mass dimension 6 (solid curve). Dimension 8 condensates are globally accumulated in a coefficient $c_4 = c_4^{(0)} + c_4^{(1)} n $~\cite{Thomas:2005dc}. The effect of such a term is exhibited by the shaded area for $|c_4^{(1)}| \leq 5 \times 10^{-5} n_0^{-1}$ GeV\,$^8$; the inclusion of $c_4^{(0)} \neq 0$ requires a readjustment of $\kappa^\mathrm{vac}_\omega$ to recover the vacuum value of $m_\omega^2$. The upper (lower) dashed curve is for $\kappa^\mathrm{med}_\omega = 3.5$ ($\kappa^\mathrm{med}_\omega = 4.5$).}
\label{fig:omegaDensity}
\end{figure}
\vspace{-0.5cm}

\paragraph{${\bar{q} q}$-sector: $\omega$ Meson}
In the photoproduction off $\mathit{Nb}$ the CB-TAPS collaboration has observed additional strength of the $\omega$ meson at lower invariant mass compared to a test reaction with a $\mathit{LH_2}$ (proton) target~\cite{Trnka:2005ey}; this is interpreted as experimental evidence for a medium-modified spectral distribution of the $\omega$ meson. One may use this finding to constrain the density dependence of a linear combination of particular four-quark condensates,
$\langle \bar u \gamma^\mu \lambda_A u \bar d \gamma_\mu \lambda_A d \rangle$,
$\langle \bar u \gamma_5 \gamma^\mu \lambda_A u \bar d \gamma_5 \gamma_\mu \lambda_A d \rangle$,
$\langle \bar q \gamma^\mu \lambda_A q \bar q \gamma_\mu \lambda_A q \rangle$ and
$\langle \bar q \gamma_5 \gamma^\mu \lambda_A q 
\bar q \gamma_5 \gamma_\mu \lambda_A q \rangle$ ($q \equiv u,d$).
For the center of the $\omega$ spectral distribution not increasing, as indicated by the data~\cite{Trnka:2005ey}, one concludes a strong decrease of the special four-quark condensate combination entering the QCD sum rule for the $\omega$ meson in the coefficient $c_3=c_3^{(0)}(\kappa^\mathrm{vac}_\omega) + c_3^{(1)} (\kappa^\mathrm{med}_\omega) n$: $\kappa^\mathrm{med}_\omega \gtrsim 4$.
The limiting situation with a constant spectral moment $m_\omega^2$ at small $n$ is depicted in Fig.~\ref{fig:omegaDensity}.

\begin{figure}[htb]
\includegraphics[width=6.9cm,angle=270]{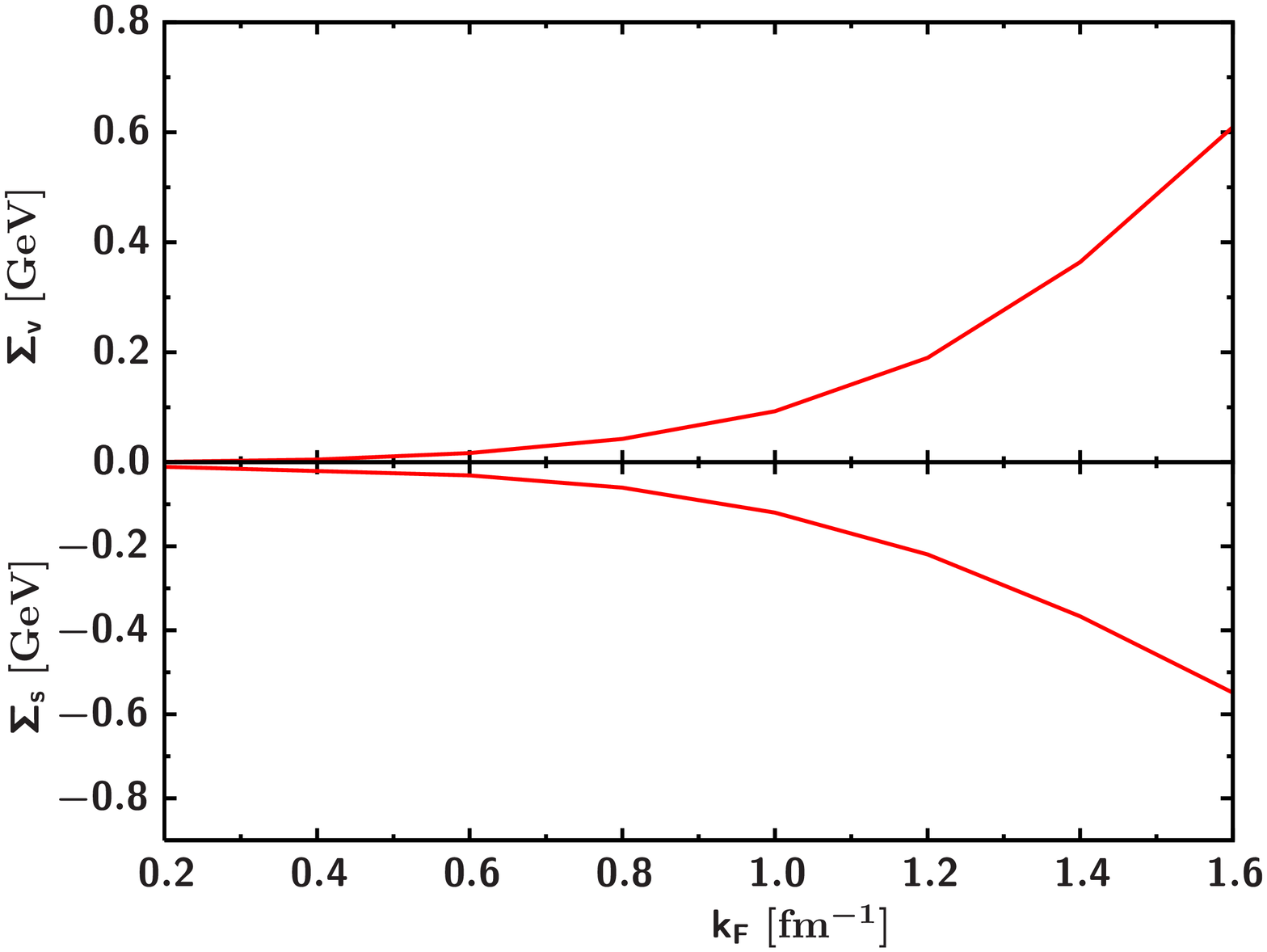}
\vspace{-0.6cm}
\caption{The scalar ($\Sigma_\mathrm{s}$) and vector ($\Sigma_\mathrm{v}$) self-energies of the nucleon with momentum $p=k_F$ at density $n(k_F)$ as functions of the Fermi momentum $k_\mathrm{F}$ for the specific choice of density dependencies of the combined four-quark condensates: $\kappa^\mathrm{med}_s =1, \kappa^\mathrm{med}_q =0, \kappa^\mathrm{med}_v =0$. Note the similarity with advanced nuclear matter calculations~\cite{Plohl:2006hy}.}
\label{fig:nucleonDensity}
\end{figure}

\paragraph{${qqq}$-sector: Nucleon}
The situation becomes more involved for the nucleon, where 3 connected sum rule equations arise~\cite{Furnstahl:1992pi}, each including another four-quark condensate combination which requires the three independent parameters $\kappa^\mathrm{med}_s$, $\kappa^\mathrm{med}_q$ and $\kappa^\mathrm{med}_v$ governing their density dependencies. Fig.~\ref{fig:nucleonDensity} exhibits qualitatively
their expected behavior reproduced in a sum rule evaluation using $\kappa^\mathrm{med}$~values derived from a perturbative chiral quark model~\cite{Drukarev:2003xd}. Especially $\kappa^\mathrm{med}_q = 0$, which corresponds to a combination remaining constant while density increases, signals a deviation from the factorization ansatz. The color structure of four-quark condensates in baryon sum rules, a mixture of both columns in Tab.~\ref{tab:fqclist}, prevents their combinations to be equated with those in the $\omega$ sum rule.

\vspace{-1.1cm}
\vspace{-0.5cm}

\section{Summary}\vspace{-0.05cm}
The research in modified hadronic properties opens access to an understanding of the QCD ground state and its response to finite density and temperature. The description of these effects via condensates can constrain particular combinations, as in the case of the $\omega$ meson or the nucleon, where four-quark condensates differently combined from an extensive catalog of possible structures tend to have either strong or weak density dependencies, respectively.
In systems containing heavy quarks, like $D$ mesons --- envisaged within the CBM and PANDA projects at FAIR --- conclusions for other dominating condensates could be drawn.

\end{document}